\documentclass[aps,prd,twocolumn,groupedaddress]{revtex4-1}
\usepackage{amsmath, amssymb}
\usepackage{multirow}
\usepackage{graphicx, caption}

\begin{document}


\title{Cosmology in a toy model of Lorentz breaking gravity} 


\author{Nils A. Nilsson}
\email[]{albin.nilsson@ncbj.gov.pl}
\affiliation{National Centre for Nuclear Research, Ho\.za 69, 00-681, Warsaw, Poland}

\author{Mariusz P. D\c{a}browski}
\email{Mariusz.Dabrowski@ncbj.gov.pl}
\affiliation{National Centre for Nuclear Research, Ho\.za 69, 00-681, Warsaw, Poland}
\affiliation{Institute of Physics, University of Szczecin, Wielkopolska 15, 70-451 Szczecin, Poland}
\affiliation{Copernicus Center for Interdisciplinary Studies, S{\l}awkowska 17, 31-016 Krak{\'o}w, Poland}


\date{\today}

\begin{abstract}
    We explore cosmological solutions to Lorentz breaking gravity using the gravitational sector of the Standard Model Extension (SME). By using a simple toy model for Lorentz violation and under the assumption that the so-called $\mathfrak{T}$-tensor is covariantly constant, we show that the gravity sector SME influences basic cosmology. If the approach used in this toy model is valid these results should point the way for more sofisticated studies.
\end{abstract}


\maketitle


\section{Introduction\label{intro}}
It is expected that known physics should break down at the Planck scale, the characteristic scale where quantum gravity effects are assumed to be important.
A systematic approach to Lorentz violation phenomenology is to use an effective field theory, and we choose to work in the Standard Model Extension (SME), which includes the standard model of particle physics, general relativity, as well as \emph{all} Lorentz and CPT violation terms~\cite{Colladay:1998fq, Kostelecky:2003fs}. These terms are constructed from an operator, which is built from the standard fields, as well as an SME coefficient which controls the magnitude of each term. As the SME contains extensions to all known physics, it is natural to divide it into sectors, and in this paper we are interested in the gravitational part of the SME. The most general way of writing this sector in the vierbein formalism is~\cite{Kostelecky:2003fs}:
\begin{equation}\label{eq:sme_grav}
\mathcal{L}_{\text{gravity}} = \underbrace{(eR-2e\Lambda + \ldots)}_{\mathcal{L}^{\text{LI}}_{e,\omega}} + \underbrace{(e(k_T)^{\lambda\mu\nu}T_{\lambda\mu\nu} + \ldots)}_{\mathcal{L}^{\text{LV}}_{e,\omega}},
\end{equation}
where $\mathcal{L}^{\text{LI}}_{e,\omega}$ is the Lorentz invariant part of the Langrangian, $\mathcal{L}^{\text{LV}}_{e,\omega}$ is the Lorentz-violating part, $e$ is the vierbein, $\omega$ is the spin connection associated with the vierbein, $R$ is the Ricci scalar, and $\Lambda$ is the cosmological constant. Moreover, $T_{\lambda\mu\nu}$ is the torsion tensor and the $k_T$ coefficients parametrise Lorentz violation. 

\section{Lorentz Violation in Gravity}
The Lagrangian for the minimal gravitational sector of the SME can be written as $\mathcal{L}_{mgSME} = k^{\mu\nu\alpha\beta}R_{\mu\nu\alpha\beta}$, where $R_{\mu\nu\alpha\beta}$ is the Riemann tensor and $k^{\mu\nu\alpha\beta}$ parametrises Lorentz violation. The term $k^{\mu\nu\alpha\beta}$ can either be thought of as a set of parameters or as an independent degree of freedom with its own dynamics. In this paper we will only consider the first case.

Through a Ricci decomposition, the action for $\mathcal{L}_{mgSME}$ can be written as:
\begin{equation}\label{eq:action}
    S_{mgSME} = \int \text{d}^4x\sqrt{-g}\left[-uR + s^{\mu\nu}R^{(T)}_{\mu\nu}+\mathfrak{T}^{\mu\nu\alpha\beta}W_{\mu\nu\alpha\beta}\right],
\end{equation}
where $R$ is the Ricci scalar, $R^{(T)}_{\mu\nu}$ is the trace-free Ricci tensor, and $W_{\mu\nu\alpha\beta}$ is the Weyl tensor. The parameters $u, s^{\mu\nu}, \mathfrak{T}^{\mu\nu\alpha\beta}$ carry the same symmetries as their accompanying tensors. It has been shown in~\cite{t-puzzle}, that $u$ and $s^{\mu\nu}$ can be removed by metric and field redefinitions, and in light of this we will use (in this toy model) the following form of the action:
\begin{equation}
    S_{mgSME} = \int \text{d}^4x\sqrt{-g}\,\mathfrak{T}^{\mu\nu\alpha\beta}W_{\mu\nu\alpha\beta}, 
    \label{SMEt} 
\end{equation}
and the total action for our theory becomes:
\begin{equation}\label{eq:main_action}
\begin{split}
    S = S_{GR} + S_M + S_\mathfrak{T} + S_{mgSME} &= \\\int \text{d}^4x\sqrt{-g}\left[R+\mathcal{L}_M+\mathcal{L}_\mathfrak{T} + \mathfrak{T}^{\mu\nu\alpha\beta}W_{\mu\nu\alpha\beta}\right],
\end{split}
\end{equation}
where $S_M=\int \text{d}^4x\sqrt{-g}\mathcal{L}_M$ is the action for any matter fields present and $S_\mathfrak{T} =\int \text{d}^4\sqrt{-g}x\mathcal{L}_\mathfrak{T}$ is the action for the tensor $\mathfrak{T}^{\mu\nu\alpha\beta}$.
The field equations obtained by varying $S$ with respect to the inverse metric $g^{\mu\nu}$ are:
\begin{equation}\label{eq:fieldeq}
\begin{split}
    T_{\mu\nu} &=
    R_{\mu\nu}-\frac{1}{2}g_{\mu\nu}R+2R^{\alpha\beta}\mathfrak{T}_{\mu\alpha\nu\beta}-R_{(\mu}^{\hspace{7pt}\alpha\beta\rho}\mathfrak{T}_{\nu)\alpha\beta\rho}-\\&\frac{1}{2}R_{\mu}^{\hspace{7pt}\alpha\beta\rho}\mathfrak{T}_{\nu\alpha\beta\rho}-R_\mu^{\hspace{7pt}\alpha\beta\rho}\mathfrak{T}_{\nu\beta\alpha\rho}-\frac{1}{2}g_{\mu\nu}\mathfrak{T}^{\alpha\beta\rho\sigma}W_{\alpha\beta\rho\sigma}+\\&2\left(\nabla_\beta\nabla_\alpha
    \mathfrak{T}_{\mu\hspace{7pt}\nu}^{\hspace{7pt}\alpha\hspace{7pt}\beta}\right).
\end{split}
\end{equation}
where $\nabla$ is the covariant derivative associated with the metric $g_{\mu\nu}$. The notation $R_{(\mu}^{\hspace{7pt}\alpha\beta\rho}\mathfrak{T}_{\nu)\alpha\beta\rho}$ implies antisymmetrisation between indices $\mu$ and $\nu$.

From now on we will restrict ourselves to a simplified scenario where the following applies: we assume that $\mathfrak{T}$ is {\it covariantly constant}, $\nabla_\lambda \mathfrak{T}^{\mu\nu\alpha\beta} = 0$. Moreover, we assume an isotropic universe, so $\partial_i \mathfrak{T}^{\mu\nu\alpha\beta} = 0$ in comoving coordinates. Moreover, as will be very important to our analysis, $\mathfrak{T}$ is completely traceless. This is not an assumption; as it appears in
the Weyl term of the the Ricci decomposition it carries all the Weyl symmetries, as was mentioned above. In this case, the field equations simplify to:
\begin{equation}\label{eq:fieldeq}
    T_{\mu\nu} = R_{\mu\nu} - \frac{1}{2}g_{\mu\nu}R -R_{\nu\alpha\beta\rho}\mathfrak{T}_\mu^{\hspace{7pt}\alpha\beta\rho}-\frac{1}{2}g_{\mu\nu}\mathfrak{T}^{\alpha\beta\rho\sigma}W_{\alpha\beta\rho\sigma},
\end{equation}
and we will use these field equations to investigate $\mathfrak{T}$-tensor Lorentz violation in two specific cosmological models.

\section{Examples of $\mathfrak{T}$- tensor effects in cosmological models}

\subsection{Friedmann-Lema\^itre-Robertson-Walker universe}
The metric of this model reads as:
\begin{equation}\label{eq:friedmann}
    ds^2 = -dt^2 + a(t)^2\left[\frac{dr^2}{1-Kr^2} + d\Omega^2\right]
\end{equation}
where $a(t)$ is the scale factor, $d\Omega$ is the solid angle element, and $K = \{-1,0,1\}$ represents a closed, flat, or open universe, respectively. 

By taking the $(0,0)$ component of the field equations~(\ref{eq:fieldeq}) we find the first Friedmann equation:
\begin{equation}\label{eq:friedmann1}
    \left(\frac{\dot{a}}{a}\right)^2 - \frac{4}{3a^2}\left(K+\dot{a}^2\right)\underbrace{\mathfrak{T}^{0\hspace{7pt}i0}_{\hspace{7pti}}}_{= 0} = \frac{8\pi G\rho}{3} - \frac{K}{a^2},
\end{equation}
so there is no apparent effect on cosmological evolution. However, by taking the trace of the spatial components of the field equations we arrive at:
\begin{equation}\label{eq:friedmann2}
    \beta_\mathfrak{T}(x^0)\frac{\ddot{a}}{a} = \frac{8\pi G}{3}\Big[3p - (\alpha_\mathfrak{T}(x^0) - 1)\rho\Big], 
\end{equation}
where $\alpha_\mathfrak{T}$ and $\beta_\mathfrak{T}$ are collections of $\mathfrak{T}$-tensor components. Since we still have some freedom in $\mathfrak{T}$, they can in principle be functions of the time coordinate $x^0$. 

In standard general relativity, it is possible to derive the corresponding Eq.~(\ref{eq:friedmann2}) from Eq.~(\ref{eq:friedmann1}) using textbook techniques. However, this derivation involves the \emph{conservation} equation $\dot{\rho} - 3(\dot{a}/a)(\rho + 3p) = 0$. In this model this relation no longer holds, as it is derived from the second Bianchi identity by demanding that the stress-energy tensor is covariantly constant, $\nabla_\mu G^{\mu\nu} = \nabla_\mu
T^{\mu\nu}=0$. This relation is modified in our model, and the second Bianchi identity now leads to the following:
\begin{equation}\label{eq:cons}
    \begin{split}
    &8\pi G\,\nabla_\mu T^{\mu\nu} + \frac{1}{2}g^{\mu\nu}\,\mathfrak{T}^{\alpha\beta\rho\sigma}\nabla_\mu W_{\alpha\beta\rho\sigma} -\\& -\nabla_\mu\left(R^{(\mu\alpha\beta\rho}\right)\,\mathfrak{T}^{\nu)}_{\hspace{7pt}\alpha\beta\rho}-2\nabla_\mu R^{\alpha\beta}\mathfrak{T}_{\alpha\beta}^{\hspace{14pt}\mu\nu} = 0.
\end{split}
\end{equation}
As a consequence of the cosmological models we have chosen to look at (FLRW and de-Sitter), this expression can be simplified further. Since both these models have a vanishing Weyl tensor and a diagonal Ricci tensor, Eq.~(\ref{eq:cons}) can be written as (setting $\nu = 0$ to obtain the correct left hand side):
\begin{equation}\label{eq:conssimple}
    8\pi G\left(\dot{\rho} - 3\frac{\dot{a}}{a}\left(\rho + 3p\right)\right) = \nabla_\mu R^{(\mu\alpha\beta\rho} \, \mathfrak{T}^{0)}_{\hspace{7pt}\alpha\beta\rho},
\end{equation}
which is the modified conservation equation in the presence of $\mathfrak{T}$-tensor Lorentz violation. The number of terms on the right hand side is greatly reduced by the tensor symmetries.This may be a path forward to gain insight in the so-called t-puzzle \cite{t-puzzle} which has been a matter of some debate in the literature~\cite{t-puzzle, Bailey:2006fd, Tasson:2012nx}. In other words we claim that, if this toy model has any merit, there is an influence of the $\mathfrak{T}$-tensor on cosmology, specifically energy
conservation. Application of the SME to inflation was investigated in~\cite{Bonder:2017dpb}.

\begin{figure}
    \begin{center}
        \includegraphics[width=0.51\textwidth]{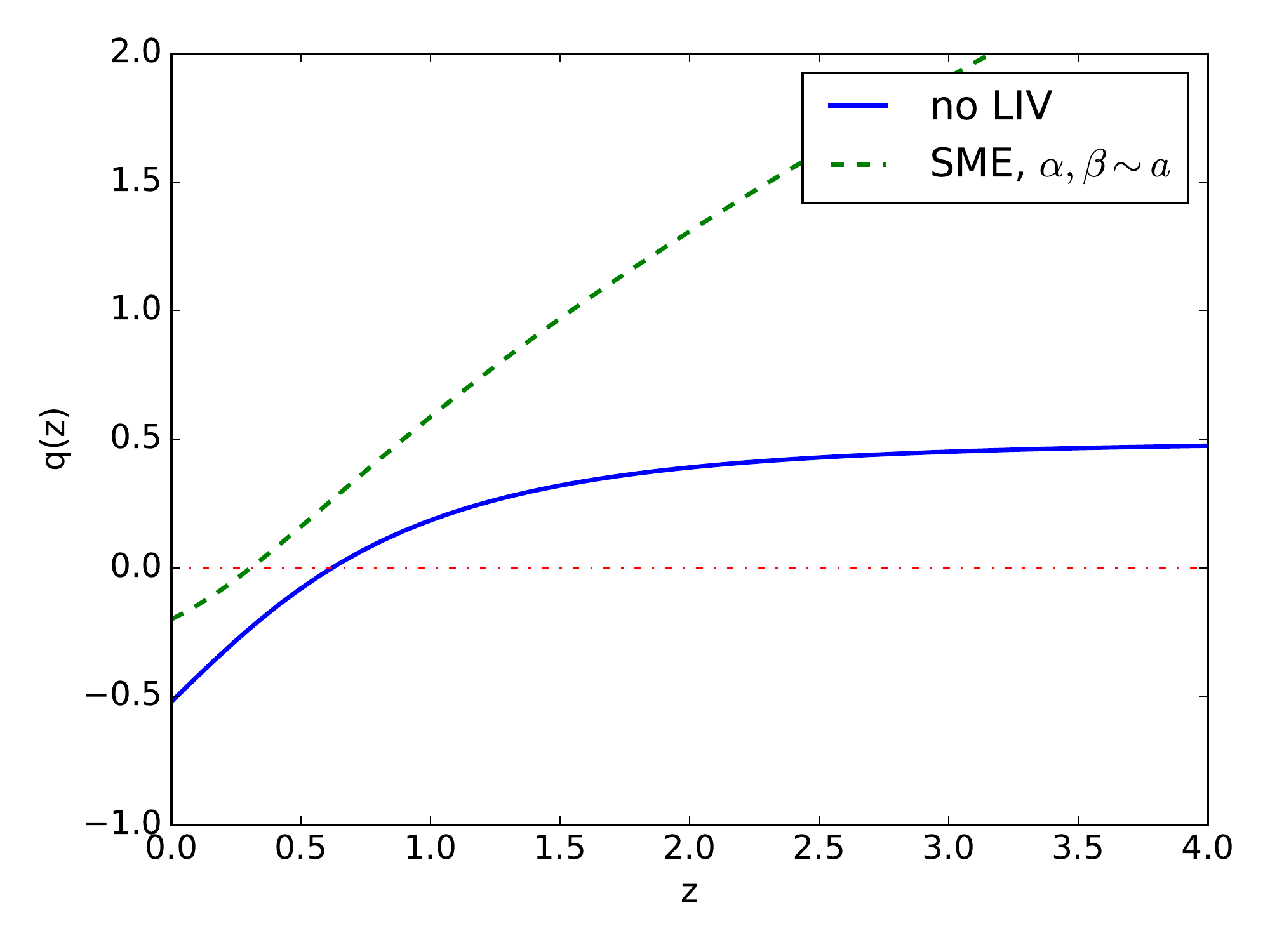}
        \caption{deceleration parameter $q$ as a function of redshift. The blue (solid) line represents standard Friedmann evolution and the green (dashed) line is our SME model. We have simplified the model by assuming that $\alpha_\mathfrak{T}$ and $\beta_\mathfrak{T}$ both evolve as the scale factor $a$.}
        \label{fig:decel}
    \end{center}
\end{figure}

As an example of possible effects from the $\mathfrak{T}$-tensor contributtions to Eq.~(\ref{eq:friedmann2}) it is useful to look at the deceleration parameter, $q = -\ddot{a}a/\dot{a}^2$, which can be constructed from the first and second Friedmann equations. Figure~\ref{fig:decel} shows the evolution of $q$ as a function of redshift $z$ for standard general relativity and for our model. In order to obtain this example we have made further simplifications to our already
straightforward toy model by assuming that the effects from Eq.~(\ref{eq:cons}) are small and that $\alpha_\mathfrak{T}$ and $\beta_\mathfrak{T}$ are proportional to the scale factor $a$. We see that the evolution of the deceleration parameter changes and may move the transition point from matter to $\Lambda$ domination. In our case we note that the dark energy domination (under some assumptions) happened later than in the standard case, but this of course requires more constraints from other cosmological tests tobe applied for a proper observational check.

\subsection{de-Sitter space}
Another spacetime worth investigating is de-Sitter space, which is obtained as a vacuum universe of constant curvature. The vacuum energy equation of state is $p_{\text{vac}} = - \rho_{\text{vac}}$.  At very early times our Universe is well approximated by de-Sitter geometry. As Lorentz violation is expected to become strong or emerge at high energies, investigating de-Sitter is of interest. 
The metric forEinstein-de-Sitter space which has flat spatial sections is:
\begin{equation}\label{eq:desitter}
    ds^2 = -dt^2 + e^{2H_{dS}t}dx_idx^i,
\end{equation}
and these coordinates $(t, \mathbf{x})$ only cover a half of de-Sitter manifold which is a one-sheet hyperboloid.
The metric (\ref{eq:desitter}) leads to the following Friedmann equation:
\begin{equation}\label{eq:dsfriedmann1}
    -H^2_{dS}\left(-1+\frac{4}{3}\underbrace{\mathfrak{T}^{0\hspace{7pt}i0}_{\hspace{7pt}i}}_{= 0}\right) = \frac{8\pi G\rho_{\text{vac}}}{3},
\end{equation}
which leads to the same expression as in general relativity, $H_{dS}^2 = 8\pi G\rho_\text{vac}/3$. However, just as in the Friedmann model, new effects appear in the trace of the spatial components of the modified field equations. In this case, however, this only leads to a constraint on some of the components of the $\mathfrak{T}$-tensor:
\begin{equation}
    \mathfrak{T}^{0\hspace{7pt}30}_{\hspace{7pt}3} = -2\mathfrak{T}^{1\hspace{14pt}1}_{\hspace{7pt}22} -\mathfrak{T}^{1\hspace{14pt}1}_{\hspace{7pt}33} - \mathfrak{T}^{2\hspace{14pt}2}_{\hspace{7pt}33}
\end{equation}
which is somewhat less illuminating than in the Friedmann case. Nevertheless, a more sofisticated treatment of $\mathfrak{T}$-tensor Lorentz violation in de-Sitter space may be interesting for the understanding of very early Universe, where energy scales may be high enough to allow for Lorentz violating effects to become strong.

\section{Discussion \& Conclusions}
In this paper we have presented simple toy model of gravity sector Lorentz violation using the effective field theory framework called the Standard Model Extension (SME). Using only the simplified case where Lorentz violation is explicitly inserted by hand into the model (mode specifically, we looked at the case where the $\mathfrak{T}$-tensor was covariantly constant and isotropic), we were able to find corrections to the second Friedmann equation, corrections stemming
from corrections to the conservation of stress-energy in such a model. Because the second Bianchi identity no longer holds on the right hand side of the modified Einstein equations it is unclear whether such a model is compatible with Riemannian geometry [REF]. Also, we have not made any assumptions regarding the covariant conservation of the stress-energy tensor itself, such as in the Brans-Dicke theory, where it is assumed to be covariantly constant, i.e. that $\nabla_{\mu} T^{\mu \nu} =0$ vanishes in (\ref{eq:cons}). As a result, particles in our model may not propagate on geodescis. As we develop this work further, we will expand our scope to include spontaneous Lorentz violation, i.e. where $\mathfrak{T}$ is a dynamical tensor field.

\acknowledgments
N.A.N. wishes to thank Viktor Svensson (Albert Einstein Institute Potsdam, NCBJ) for valuable discussions. M.P.D. acknowledges the discussions with Alan Kostelecky and Ralf Lehnert. This work was financed by the Polish National Science Center Grant DEC-2012/06/A/ST2/00395. Parts of this note was presented as a poster at the \emph{Third IUCSS Summer School on the Lorentz and CPT violation Standard Model Extension (SME)}, June 14-23 2018.

\bibliography{mybib.bib}

\end{document}